\documentclass[aps,preprint,superscriptaddress, showpacs, showkeys]{revtex4}

\usepackage{graphicx}

\newcommand{\be}{\begin{equation}}
\newcommand{\ee}{\end{equation}}
\newcommand{\bea}{\begin{eqnarray}}
\newcommand{\eea}{\end{eqnarray}}

\newcommand{\vfi}{\varphi}
\newcommand{\Vm}{V^-}

\newcommand{\Fiu}{\Phi^u}
\newcommand{\Fip}{\Phi^p}

\newcommand{\LL}{\frac{1}{2\pi}\,\int_{-\infty}^{\infty}dk\sum_{n=-\infty}^{\infty}}
\newcommand{\LLnof}{\int_{-\infty}^{\infty}dk\sum_{n=-\infty}^{\infty}}
\newcommand{\FF}{\frac{1}{2\pi}\,\int\limits_{0}^{2\pi}\int\limits_{-\infty}^{\infty}d\vfi dz}

\begin{document}

\title{\vspace{0cm}\large\bf
Electrostatic Patch Effect in Cylindrical Geometry. I. \\
Potential and Energy between Slightly Non--Coaxial Cylinders}

\author{Valerio Ferroni}
\affiliation{ICRANet, Dept. of Phys., Univ.  `La Sapienza', Rome, Italy\\
{\rm current address}: W.W.Hansen Experimental Physics Laboratory,\\
Stanford University, Stanford, CA 94305-4085, USA}
\email{vferroni@stanford.edu}
\author{Alexander S. Silbergleit}
\affiliation{Gravity Probe B, W.W.Hansen Experimental Physics Laboratory,\\
Stanford University, Stanford, CA 94305-4085, USA}
\email{gleit@stanford.edu}

\date{\today}

\begin{abstract}
We study the effect of any uneven voltage distribution on two close cylindrical conductors with parallel axes that are slightly shifted in the radial and by any length in the axial direction. The investigation is especially motivated by certain precision measurements, such as the Satellite Test of the Equivalence Principle (STEP).

By energy conservation, the force can be found as the energy gradient in the vector of the shift, which requires determining potential distribution and energy in the gap. The boundary value problem for the potential is solved, and energy is thus found to the second order in the small transverse shift, and to lowest order in the gap to cylinder radius ratio.  The energy consists of three parts: the usual capacitor part due to the uniform potential difference, the one coming from the interaction between the voltage patches and the uniform voltage difference, and the energy of patch interaction, entirely independent of the uniform voltage. Patch effect forces and torques in the cylindrical configuration are derived and analyzed in the next two parts of this work. 
\end{abstract}


\maketitle

\section{Introduction\label{s1}}

\subsection{Background and Motivation\label{s1.1}}

The electrostatic patch effect (PE)~\cite{Dar} is a nonuniform potential distribution on the surface of a metal. Charges in a metal move, due to a finite conductivity, until the electrostatic potential becomes the same at all the surface points. However, a nonuniform dipole layer may form on the metal surface due to impurities or microcrystal structure. If it exists, then the potential on the surface is no longer uniform, and the electric field is not necessarily perpendicular to the surface of the metal. In the (idealized) case of an isolated conductor the net force and torque on the body is still equal to zero, but with two metallic surfaces at a finite distance, the net force and/or torque on each of them does not vanish, in general. This effect is larger, the closer the surfaces, as first confirmed by the calculation of the patch effect force for two parallel conducting planes~\cite{Sp}. 

This calculation of the patch effect force was particularly motivated by the LISA space experiment to detect gravitational waves (see c.f.~\cite{LISA}). 
PE can similarly affect the accuracy of any other precision measurement if its set-up includes conducting surfaces in a closed proximity to each other. For instance, PE torques turned out one of the two major difficulties~\cite{E, H&S, BT} in the analysis of data from Gravity Probe B (GP-B) Relativity Science Mission; the GP-B satellite was in flight in 2004--2005, to measure the relativistic drift of a gyroscope predicted by Einstein's general relativity (see c.f.~\cite{KetAl}). This required theoretical calculation of PE torques~\cite{KKS} for the case of two concentric spherical conductors (the lab evidence of the patch formation on the surfaces of the rotor and housing was found in~\cite{BG}).

The aim of this paper, consisting of three parts, is to study the force and torque due to voltage patches on two close cylindrical conductors. In particular, in the STEP experimental configuration~\cite{PW, Mes, PWMes, Over} each test mass and its superconducting magnetic bearing form such a pair of cylinders. The goal of STEP is the precise (1 part in $10^{18}$) measurement of the relative axial acceleration of a pair of coaxial test masses, so the importance of properly examining cylindrical patch effect is evident.

\subsection{Structure of the Paper\label{s1.2}}

We determine the PE forces (and axial torque, in the final part of this paper) between two cylinders with parallel axes by the energy conservation argument. It implies, in general, that a small shift, $\vec{r}{\;^0}$, of one of the conductors relative to the other causes an electrostatic force between them given by (see, for instance,~\cite{Smy})
\be
\vec F(\vec{r}^{\;0}) = -\frac{\partial W(\vec{r}^{\;0})}{\partial\vec{r}^{\;0}} \; ,
\label{forcegen}
\ee
where $W(\vec{r}{\;^0})$ is the electrostatic energy as a function of the shift. 

According to the formula (\ref{forcegen}), the force to linear order in a small shift, which is our ultimate goal, requires calculating $W(\vec{r}{\;^0)}$ to quadratic terms in $\vec{r}{\;^0}$. Due to the specifics of cylindrical geometry, a more general result for an arbitrary axial shift, $z^0$, but only small transverse shifts, $x^0=x_1^0,\;y^0=x_2^0$ is available. So, we first find the energy in the form
\be
W(\vec{r}{\;^0})=W_0(z^0)+W_\mu(z^0)\,x^0_\mu+\;W_{\mu\nu}(z^0)\,x^0_\mu\,x^0_\nu+O\left(\rho_0^3\right),\quad
\rho_0\equiv\sqrt{\left(x^0\right)^2+\left(y^0\right)^2} \; .
\label{engen}
\ee
Here and elsewhere in the paper we use the summation rule over repeated Greek indices: the summation always runs from 1 to 2 (meaning the directions transverse to the cylinder axes). To compute the coefficients $W_0(z^0),\;W_\mu(z^0),\;W_{\mu\nu}(z^0)$, and then the force components by the formulas (\ref{forcegen}), we need the distribution of the electrostatic potential in the gap between the shifted cylinders to the quadratic order in the shift, 
\be
\Phi(\vec{r},\vec{r}{\;^0})=\Phi_0(\vec{r},z^0)+\Phi_\mu(\vec{r},z^0)\,x^0_\mu+\Phi_{\mu\nu}(\vec{r},z^0)\,x^0_\mu\,x^0_\nu+
O\left(\rho_0^3\right) \; ,
\label{Figen}
\ee
since energy is a quadratic functional of the potential.

A more rigorous view of expansions in the above formulas consists of assuming the transverse shift, $\rho_0$, small as compared to the nominal gap, $d=b-a$, between the coaxial cylinders ($a<b$ are the inner and outer cylinder radii). The relevant small parameter is $\rho_0/d$, and the expansions go in dimensionless quantities $x^0_\mu/d,\;\mu=1,2$, so the remainder estimate $O\left(\rho_0^3\right)$ means, in fact, $O\left(\rho_0^3/d^3\right)$, etc. 
Also, for typical experimental conditions, such as the STEP configuration ~\cite{Mes, Over}, the gap is much smaller than either of the radii. This justifies the model of infinite cylinders, and allows for a significant simplification of the results to a certain order in $d/a$. Thus two small parameters are actually involved in the problem,
\be
\rho_0/d\,\ll\,1,\qquad  d/a\,\sim \,d/b\,\ll\,1 \; .
\label{smalpar}
\ee
We work all the way to quadratic order in the first of them, and give the final answer for the PE forces to lowest order in the second one. However, some meaningful intermediate results are valid without the last or both of these assumptions.

All said pretty much defines the structure of the paper. In the next section the boundary value problem (BVP) for the potential with general voltage distributions on the cylinders is solved, and the potential in the form (\ref{Figen}) is found. Based on this, the energy representation (\ref{engen}) is obtained explicitly in section \ref{s3}. The details of calculations, in places rather complicated and cumbersome, are found in the three appendices. The derivation of PE forces and torques and the analysis of their properties are given in the next two parts of this work. The results of this part can also be used whenever the solution to the BVP for the Laplace equation for the domain between two cylinders is needed, such as for instance, in magnetostatics, thermostatics, stationary diffusion, etc.

\section{Electrostatic Potential Between Two \\Infinite Cylinders with Parallel Axes\label{s2}}

\subsection{Boundary Value Problem\label{s2.1}}

\begin{figure}[ht]
\centering
\includegraphics[scale=0.5]{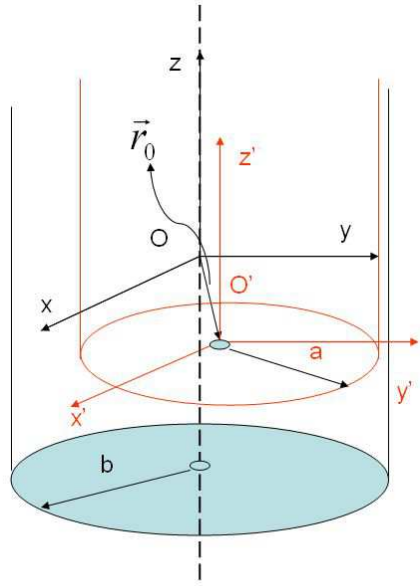}
\caption{{Geometry of the problem and coordinate systems}}
\label{fig1}
\end{figure}

We employ both Cartesian and cylindrical coordinates in two frames of the inner and outer cylinders as shown in fig.\ref{fig1}. In the inner, or `primed', frame the position of a point is given by the vector radius $\vec{r}\;^{'}$, and Cartesian coordinates $\{x^{\;'},y^{\;'},z^{\;'}\}$ or cylindrical coordinates $\{\rho^{\;'},\vfi^{\;'},z^{\;'}\}$. In the outer, or `unprimed', frame the corresponding quantities are $\vec{r}$, $\{x,y,z\}$,  $\{\rho,\vfi,z\}$. The frame origins are separated by $\vec{r}{\;^0}$, hence the primed and unprimed Cartesian coordinates are related by
\be
\vec{r}\;^{'} = \vec{r}+\vec{r}^{\;0};\qquad 
x\;^{'} = x+x^{\;0},\quad y\;^{'} = y+y^{\;0}, \quad z\;^{'} = z+z^{\;0}\; ;
\label{coord_shift}
\ee
equivalenlty, in cylindrical components of the transverse shift [$\rho_0$ is defined in (\ref{engen})],
\be
x\;^{'} = x+\rho_0\cos\vfi_0,\quad y\;^{'} = y+\rho_0\sin\vfi_0; \qquad \tan\vfi_0 = y^{\;0}/x^{\;0}\; ,
\label{cyl_shift}
\ee
As alternative writing we use $x^{\;0}\equiv x_1^{\;0},\; y^{\;0}\equiv x_2^{\;0},\; x\equiv x_1,\; y\;^{'}\equiv x_2^{'}$, etc.

The surfaces of the inner and outer cylinders are $\rho^{\;'}=a$ and $\rho=b$, respectively. They carry arbitrary voltage distributions, so the electrostatic potential, $\Phi$, satisfies the Laplace equation in the gap,
\be
\Delta\Phi =0,\qquad \rho^{\;'}>a,\quad \rho<b,\quad 0\leq\vfi<2\pi,\quad |z|<\infty\; ,
\label{Lapl}
\ee 
and the boundary conditions of the first kind at the boundaries:
\bea
\Phi\biggl|_{\rho^{\;'}=a}=G(\vfi^{\;'},z^{\;'}),\qquad \Phi\biggl|_{\rho=b}\;\;=\Vm+H(\vfi,z)\; ;\label{bca}\\
\Vm=const\; \qquad\qquad\qquad\qquad\qquad 
\label{Vu}
\eea
the latter are formulated using two sets of coordinates, primed  and unprimed. [The potential distribution can also be described by using either of them, such as $\Phi(\vec{r}\;^{'})\equiv\Phi(\vec{r}\;^{'}, \vec{r}\;^{0})$, or $\Phi(\vec{r})\equiv\Phi(\vec{r}, \vec{r}\;^{0})$; the second argument emphasizes the dependence on the shift between the cylinders.] Boundary voltages are split in two parts: the uniform potential difference, $\Vm$ (all the voltages in the problem are counted from the uniform voltage of the inner cylinder taken as zero), and the non--uniform potential distributions (patch patterns) described by arbitrary smooth enough functions $G(\vfi^{\;'},z^{\;'})$ and $H(\vfi,z)$.

The local nature of the patch distributions is stressed by requiring
\be
||G||^2=\int\limits_{0}^{2\pi}\int\limits_{-\infty}^{\infty}d\vfi^{\;'}dz^{\;'}\,|G(\vfi^{\;'},z^{\;'})|^2<\infty,\qquad
||H||^2=\int\limits_{0}^{2\pi}\int\limits_{-\infty}^{\infty}d\vfi dz\,|H(\vfi,z)|^2<\infty\; ;
\label{L2}
\ee
these conditions are assumed valid throughout the paper. (Of course, primes can be dropped at the variables under the integrals, as done everywhere below). Later we will assume the boundary functions more smooth, expressing the additional conditions as the square integrability of various derivatives of $G(\vfi^{'},z^{'})$ and $H(\vfi,z)$.

The main tool in the solution of the boundary value problem (BVP) (\ref{Lapl}), (\ref{bca}) is the Fourier transform in the axial and azimuthal variables. For any function $u(\vfi,z)$ satisfying the square integrability condition the following representations hold:
\bea
u(\vfi,z)=\LL\,u_n(k)e^{i(kz+n\vfi)},\quad u_n(k)=\FF\,u(\vfi,z)e^{-i(kz+n\vfi)}. 
\label{Four}
\eea
For any two such functions $u(\vfi,z)$ and $v(\vfi,z)$ the useful Parceval identity holds:
\be
(u,\,v)\equiv\int\limits_{0}^{2\pi}\int\limits_{-\infty}^{\infty}d\vfi dz\,u(\vfi,z)v^*(\vfi,z)= 
\LLnof u_n(k)v^*_n(k)\; ;
\label{Parc}
\ee
here and elsewhere the star denotes complex conjugation. In a particular case $u=v$ the identity (\ref{Parc}) shows that the square of the norm, $||u||^2$, of a function $u$, defined by the formula (\ref{L2}), is equal to the square of the norm of its Fourier coefficient $u_n(k)$.

Finally, we note that the split in the boundary conditions  implies the corresponding representation of the potential,
\be
\Phi(\vec r)\; =\; \Fiu(\vec r)\;+\;\Fip(\vec r)\; ,
\label{Fiup}
\ee 
where $\Fiu$ is originated by the uniform boundary voltage, and $\Fip$ by the patch one.

\subsection{Solution for the  Patch Potential\label{s2.2}}

According to the formulas (\ref{Lapl}), (\ref{bca}) and (\ref{Fiup}), the BVP for the patch potential is:
\be
\Delta\Fip =0,\qquad \rho^{\;'}>a,\quad \rho<b,\quad 0\leq\vfi<2\pi,\quad |z|<\infty\; ;
\label{Laplp}
\ee
\bea
\Fip\biggl|_{\rho^{\;'}=a}=G(\vfi^{\;'},z^{\;'})\;=\;\frac{1}{2\pi}\,\LLnof{G_n(k)}e^{i(kz^{\;'}+n\vfi^{\;'})}\; ;\label{bcap}\\
\Fip\biggl|_{\rho=b}\;\;=H(\vfi,z)\;=\;\frac{1}{2\pi}\,\LLnof{H_n(k)}e^{i(kz+n\vfi)}\; .
\label{bcbp}
\eea 
Since the boundary functions are real, their Fourier coefficients satisfy important relations,
\be
G_n(k)=G^*_{-n}(-k),\qquad H_n(k)=H^*_{-n}(-k)\; ;
\label{symmGnHn}
\ee
which are necessary and sufficient conditions for the corresponding imaginary parts to vanish. The dimension of functions $G_n(k)$ and $H_n(k)$ is $volt\cdot meter$, since $k$ is in $inverse\;\; meters$. 

The standard separation of variables in the Laplace equation in cylindrical coordinates (see c.f.~\cite{Leb}, Chs. 5, 6) provides the following representation of the potential:
\be
\Fip(\vec{r}\;^{'})=\LL\left[A_n(k)\,I_n(k\rho^{\;'})+B_n(k)\,K_n(k\rho^{\;'})\right]e^{i(kz^{\;'}+n\vfi^{\;'})}\; ,
\label{preprpr}
\ee 
where $I_n(\xi),\;K_n(\xi)$ are the modified Bessel functions of the 1st and 2nd kind, respectively. In fact, the Macdonald's function $K_n(\xi)$ requires some definition for the negative values of its argument, which is taken here according to the parity of $I_n(\xi)$: the symbol $K_n(k\rho^{\;'})$ stands actually for $(\mbox{sign}\;k)^n K_n(|k|\rho^{\;'})$.

The unknowns $A_n(k),\;B_n(k)$ are to be determined from the boundary conditions; the first of them, condition (\ref{bcap}) at the inner cylinder, can apparently be fulfilled immediately:
\be
A_n(k)\,I_n(ka)\;+\;B_n(k)\,K_n(ka)\;=\;G_n(k),\qquad n=0,\pm1,\pm2,\ldots\; .
\label{1stbc}
\ee
To satisfy the remaining boundary condition  (\ref{bcbp}) we express the basis solutions of the Laplace equation in the primed cylindrical coordinates, $I_n(k\rho^{\;'})e^{i(kz^{\;'}+n\vfi^{\;'})},\;K_n(k\rho^{\;'})e^{i(kz^{\;'}+n\vfi^{\;'})}$, in terms of the set of same solutions in the unprimed coordinates of the outer cylinder. This re--expansion approach has been successfully used for solving various BVPs with the basis solutions in spherical coordinates~\cite{HH},~\cite{GS2}, and in spherical and cylindrical coordinates~\cite{HH},~\cite{GS1}. The needed re-expansions of cylindrical solutions are readily available from~\cite{Ye}; the corresponding transformation formulas, along with all the detailed calculations, are found in Appendix A. In particular, the potential is:
\bea
\Fip(\vec{r})&=&\LL\left[\tilde{A}_n(k)\,I_n(k\rho)+\tilde{B}_n(k)\,K_n(k\rho)\right]e^{i(kz+n\vfi)}\; ;\label{preprunpr}\\
\tilde{A}_n(k)&=&e^{ikz^0}\sum\limits_{m=-\infty}^{\infty}\,A_m(k)\,I_{m-n}(k\rho_0)\,e^{i(m-n)\vfi_0}
\; ;\label{tildAn}\\
\tilde{B}_n(k)&=&e^{ikz^0}\sum\limits_{m=-\infty}^{\infty}\,B_m(k)(-1)^{m-n}I_{m-n}(k\rho_0)\,e^{i(m-n)\vfi_0}
\; ;\label{tildBn}\\
& & n,\,m=0,\pm1,\pm2,\ldots\; .\qquad\qquad\qquad\nonumber
\eea 

The boundary condition (\ref{bcbp}) at the outer cylinder can now be satisfied by requiring
\be
\tilde{A}_n(k)\,I_n(kb)\;+\;\tilde{B}_n(k)\,K_n(kb)\;=\;H_n(k),\qquad n=0,\pm1,\pm2,\ldots\; .
\label{2ndbc}
\ee
Written explicitly using the expressions (\ref{tildAn}), (\ref{tildBn}), this, along with the equation (\ref{1stbc}), gives an infinite system of linear algebraic equations for the coefficients $A_n(k)$ and  $B_n(k)$:
\bea
I_n(ka)\,A_n(k)\;+\;K_n(ka)\,B_n(k)\;=\;G_n(k),\;\;\quad n=0,\pm1,\pm2,\ldots\; ;
\nonumber\\
I_n(kb)\sum\limits_{m=-\infty}^{\infty}\,A_m(k)\,I_{m-n}(k\rho_0)\,e^{i(m-n)\vfi_0}
\,+\qquad\quad\qquad\qquad\qquad\quad\label{infsys}\\
K_n(kb)\,\sum\limits_{m=-\infty}^{\infty}\,B_m(k)(-1)^{m-n}\,I_{m-n}(k\rho_0)\,e^{i(m-n)\vfi_0}
=H_n(k)e^{-ikz^0}\; .\nonumber
\eea
The solution to this system provides the patch potential for any values of the parameters involved. However, all we need is a perturbative 2nd--order solution in a small transverse shift [recall the condition (\ref{smalpar}) and our summation convention]:
\bea
A_n(k)=A^0_n(k)+A^\mu_n(k)\,x^0_\mu+A^{\mu\nu}_n(k)\,x^0_\mu\,x^0_\nu+
O\left(\rho_0^3\right)\; ;\qquad\qquad\qquad\qquad\qquad\;\;\label{AnBn}\\
B_n(k)=B^0_n(k)+B^\mu_n(k)\,x^0_\mu+B^{\mu\nu}_n(k)\,x^0_\mu\,x^0_\nu+
O\left(\rho_0^3\right)\;,\qquad n=0,\pm1,\pm2,\ldots\; \nonumber.
\eea
The system (\ref{infsys}) is perfect for these perturbations: all its non-diagonal matrix elements are small due to the factor $I_{m-n}(k\rho_0)\sim O\left(\rho_0^{|m-n|}\right)$. The calculations are done, and the expressions for all the coefficients are determined in Appendix A. They are subsequently simplified there to l. o. in the second small parameter from (\ref{smalpar}), $d/a\ll 1$, allowing for the expansion of the patch potential in both primed and unprimed coordinates in the form:
\be
\Fip=\Fip_0+\Fip_\mu\,(x^0_\mu/d)+\Fip_{\mu\nu}\,(x^0_\mu/d)\,(x^0_\nu/d)+
O\left((\rho_0/d)^3\right) \; .
\label{Fipexp}
\ee
To save space, we introduce a special notation
\be
{\cal F}\equiv\LL\; ;\label{NotFour}
\ee
using it, in the inner cylinder coordinates, to l. o. in $d/a\ll 1$, we have:
\be
\Fip_0(\vec{r}\;^{'})=-\frac{a}{d}\,{\cal F}
\left\{
\left[G_n(k)-H_n(k)e^{-\imath\,kz^0}\right]\Omega_n(k\rho^{\;'})e^{\imath\left(kz^{\;'}+n\vfi^{\;'}\right)}\right\}\, ;\label{explFip0}\quad
\ee
\be
\Fip_\mu(\vec{r}\;^{'})=\frac{a}{d}\,{\cal F}
\left\{
c_\mu^\pm\left[G_{n\pm\,1}(k)-H_{n\pm\,1}(k)e^{-\imath\,kz^0}\right]\Omega_n(k\rho^{\;'})e^{\imath\left(kz^{\;'}+n\vfi^{\;'}\right)}
\right\}\label{explFipmu}\, ;\qquad\qquad\qquad
\ee
\bea
\Fip_{\mu\nu}(\vec{r}\;^{'})=-\frac{a}{d}{\cal F}
\left\{
\left[
c_\mu^\pm c_\nu^\pm\left(G_{n\pm2}(k)-H_{n\pm2}(k)e^{-\imath\,kz^0}\right)+\right.\right.\qquad\qquad\qquad\qquad\qquad\qquad\;\;\;\nonumber\\
\left.\left.\delta_{\mu\nu}/2\left(G_n(k)-H_n(k)e^{-\imath kz^0}\right)\right]\Omega_n(k\rho^{\;'})e^{\imath\left(kz^{\;'}+n\vfi^{\;'}\right)}
\right\};\qquad\quad\mu,\,\nu=1,2\;  \label{explFipmunu}\;\;\;\;
\eea 
[see formulas (A.15)---(A.17) of Appendix A]. Here
\be
\Omega_n(k\rho^{\;'})=K_n(kb)I_n(k\rho^{\;'})-I_n(ka)K_n(k\rho^{\;'})\, ,
\label{Omega}
\ee
and terms like $c^\pm\,G_{n\pm\,1}$ and $c^\pm c^\pm\,G_{n\pm\,1}$ should be read as the sums
\[
c^\pm\,G_{n\pm\,1}=c^+\,G_{n+1}+c^-\,G_{n-1}\,, \qquad c^\pm c^\pm\,G_{n\pm\,1}=c^+c^+\,G_{n+1}+c^-c^-\,G_{n-1}\, ;
\]
(the definitions of $c^{\pm}_\mu$ are given in Appendix A). For $|k|\to \infty$, the large--argument asymptotics of the Bessel functions (\cite{Leb}, Ch.~5) leads to 
\be
\Omega_n(k\rho^{\;'})\sim |k|^{-1/2}\left[e^{-|k|(b-\rho^{\;'})}-e^{-|k|(\rho^{\;'}-a)}\right]\; ,
\label{largek}
\ee
so the integrals converge and the above representations hold for $a<\rho^{\;'}<b$.

For brevity, we omitted the factor $[1+O(d/a)]$ that expresses the correction in terms of our small parameter $d/a$ in all the formulas (\ref{explFip0})---(\ref{explFipmunu}). The estimates of the remainder are only valid under some additional, as compared to (\ref{L2}), smoothness conditions on $G(\vfi^{\;'},z^{\;'}),\;H(\vfi,z)$, namely, the square integrability of their mixed second derivatives [see formulas (\ref{L2d}), (\ref{L2der}), Appendix A]. 

By a similar fashion, we obtain the potential in the coordinates of the outer cylinder, as explained at the end of Appendix A:
\be
\label{uexplFip0}
\Fip_0(\vec{r})=-\frac{a}{d}{\cal F}
\left\{
\left[G_n(k)e^{\imath\,kz^0}-H_n(k)\right]\Omega_n(k\rho)e^{\imath\left(kz+n\vfi\right)}
\right\}\; ;\qquad\qquad\qquad\qquad\qquad\qquad\;\;\;
\ee
\be
\label{uexplFipmu}
\Fip_\mu(\vec{r})=\frac{a}{d}{\cal F}
\left\{
c_\mu^\pm\left[G_{n\pm\,1}(k)e^{\imath\,kz^0}-H_{n\pm\,1}(k)\right]\Omega_n(k\rho)e^{\imath\left(kz+n\vfi\right)}
\right\}\; ;\qquad\qquad\qquad\qquad\qquad\;
\ee
\bea
\label{uexplFipmunu}
\Fip_{\mu\nu}(\vec{r})=-\frac{a}{d}{\cal F}\left\{
\left[
c_\mu^\pm\,c_\nu^\pm\left(G_{n\pm\,2}(k)e^{\imath\,kz^0}-H_{n\pm\,2}(k)\right)+\delta_{\mu\,\nu}/2\left(G_n(k)e^{\imath\,kz^0}-H_n(k)\right)
\right]\times\right.\nonumber\qquad\\
\left.\Omega_n(k\rho)e^{\imath\left(kz+n\vfi\right)}\right\}\,,\,\qquad\qquad\mu,\,\nu=1,2\; , \qquad\qquad\qquad\qquad\qquad\qquad\qquad
\eea 
with $\Omega_n(k\rho)$ defined in (\ref{Omega}), and the same meaning of the terms $c^\pm\,G_{n\pm\,1}$, etc. Once again we dropped the factor $[1+O(d/a)]$, which is true under the same conditions (\ref{L2d}), (\ref{L2der}). Using the large argument asymptotics of the Bessel functions again, one shows the formulas (\ref{uexplFip0})--(\ref{uexplFipmunu}) to hold in the domain $a<\rho<b$, so that the combination of the two representations, in the unprimed and primed coordinates, covers the whole domain of the gap between the cylinders, that is, $a<\rho^{\;'},\;\rho<b$. 

\subsection{Alternative Method of Finding the Patch Potential\label{s2.3}}

The solution for the patch potential can also be obtained by the method of perturbation of the boundary~\cite{Nay} exploiting the first of the parameter conditions (\ref{smalpar}). Indeed, by the coordinate transformation (\ref{coord_shift}), using $(\rho_0/\rho)\sim(\rho_0/a)\ll 1$, one finds
\[
\rho\;^{'}=\sqrt{\rho^2+\rho_0^2+2\rho\rho_0\cos(\vfi-\vfi_0)}=\rho+(\rho_0/\rho)\,\cos(\vfi-\vfi_0)+\ldots\; .
\]
Hence the boundary equation $\rho\;^{'}=a$ is equivalent to
\bea
\rho=a-\epsilon(\vfi)\; ,\qquad\qquad\qquad\label{roa}\\
\epsilon(\vfi)=a_\mu(\vfi)\,x^0_\mu+b_{\mu\nu}(\vfi)\,x^0_\mu\,x^0_\nu+O\left(\rho_0^3\right) \; ,
\label{eps}
\eea
with the coefficients $a_\mu,\;b_{\mu\nu}$ found explicitly. Expanding the l.h.s of the boundary condition~(\ref{bcap}) at the inner cylinder  in a Taylor series according to the equality (\ref{roa}) we find:
\be
\Fip\biggl|_{\rho\;^{'}=a}=\Fip\biggl|_{\rho=a-\epsilon}=\Fip\biggl|_{\rho=a}-\;\frac{\partial\Fip}{\partial\rho}\biggl|_{\rho=a}\,\epsilon+
\frac{1}{2}\,\frac{\partial^2\Fip}{\partial\rho^2}\biggl|_{\rho=a}\,\epsilon^2+
O\left(\rho_0^3\right)=G_n(k) \; .
\label{bcexp}
\ee
The potential and the boundary value function can be written in the form
\bea
\Fip(\vec{r})=\Fip_0(\vec{r})+\Fip_\mu(\vec{r})\,x^0_\mu+\Fip_{\mu\nu}(\vec{r})\,x^0_\mu\,x^0_\nu+O\left(\rho_0^3\right) \; ,
\label{Fipexpunpr}\\
G(\vfi^{\;'},z^{\;'})=G_0(\vfi,z)+G_\mu(\vfi,z)\,x^0_\mu+G_{\mu\nu}(\vfi,z)\,x^0_\mu\,x^0_\nu+O\left(\rho_0^3\right) \; .
\label{gexpunpr}
\eea
Introducing the expansions (\ref{eps}), (\ref{Fipexpunpr}) and (\ref{gexpunpr}) to the boundary condition (\ref{bcexp}) and equating the coefficients at the same order on either side, we arrive at the set of the inner boundary conditions for the potentials of all orders:
\bea
\Fip_0\Biggl|_{\rho=a}=G_0(\vfi,z);\quad \Fip_\mu\Biggl|_{\rho=a}=G_\mu(\vfi,z)+\frac{\partial\Fip_0}{\partial\rho}
\Biggl|_{\rho=a}\,a_\mu(\vfi),\qquad\mu,\,\nu=1,2\; ;\nonumber\\
\label{bcaseq}
\Fip_{\mu\nu}\Biggl|_{\rho=a}=G_{\mu\nu}(\vfi,z)+\left[
\;\frac{\partial\Fip_\mu}{\partial\rho}\,a_\nu(\vfi)+\frac{\partial\Fip_0}{\partial\rho}\,b_{\mu\nu}(\vfi)-
\frac{1}{2}\,\frac{\partial^2\Fip_0}{\partial\rho^2}\,a_{\mu}(\vfi)a_{\nu}(\vfi)
\right]\Biggl|_{\rho=a}\quad\;\nonumber\; .
\eea
These potentials satisfy also the following boundary conditions at the outer cylinder, as implied by (\ref{bcbp}) and (\ref{Fipexpunpr}):
\be
\Fip_0\biggl|_{\rho=b}=H(\vfi,z);\qquad
\Fip_\mu\biggl|_{\rho=b}=\;\Fip_{\mu\nu}\biggl|_{\rho=b}=0,\qquad\mu,\,\nu=1,2\; ,
\label{bcbseq}
\ee
and of course they are solutions to the Laplace equation in the gap $a<\rho<b$. 

Thus we have a sequence of boundary value problems that can be solved by the standard separation of variables in the unprimed cylindrical coordinates, obtaining the potential to the required order. This approach turns out much more cumbersome than that of the previous section. Nevertheless, it provides an important cross--check; for this reason we have carried it out and eventually obtained exactly the same result for the patch potential.

\subsection{Potential Due to Uniform Voltages and the Final Result\label{s2.4}}

It remains to determine the potential, $\Fiu$, generated by the uniform boundary voltages. As seen from the equations (\ref{Lapl}), (\ref{bca}) and representation (\ref{Fiup}), the BVP for it is:
\bea
\Delta\Fiu =0,\qquad \rho\;^{'}>a,\quad \rho<b,\quad 0\leq\vfi<2\pi,\quad |z|<\infty\; ;\label{Laplu} \\
\Fiu\biggl|_{\rho\;^{'}=a}=0,\qquad\qquad
\Fiu\biggl|_{\rho=b}\;\;=\Vm\; ,\qquad\qquad\qquad\nonumber
\eea
where $\Vm$ is the given constant voltage from the formula (\ref{Vu}). This problem is a two--dimensional one, since the uniform potential does not depend on the axial coordinate. Its solution can be obtained by the method of section ~\ref{s2.2} without even using the re-expansions of the three--dimensional cylindrical solutions. However, formally the BVP (\ref{Laplu}) is a particular case of the BVP\break (\ref{Lapl}), (\ref{bca}) in which the boundary data are specified through their Fourier coefficients as
\be
G_n(k)=0,\qquad\qquad H_n(k)=2\pi\Vm\,\delta(k)\,\delta_{n0}\; ,
\label{gnhnu}
\ee
where $\delta(k)$ and $\delta_{n0}$ are the Dirac delta-function and the Kronecker symbol, respectively. With certain caution, the solution can be thus obtained from the results of section \ref{s2.2}, namely, formulas (\ref{explFip0})---(\ref{explFipmunu}) and (\ref{uexplFip0})---(\ref{uexplFipmunu}). The appropriate calculations, with some physical explanations, are found in Appendix B; here we give just the resulting expressions. 

The solution of the above BVP in the gap $a<\rho^{\;'},\;\rho<b $ is of the form:
\be
\Fiu=\Fiu_0+\Fiu_\mu\,\left(x^0_\mu/d\right)+\Fiu_{\mu\nu}\,\left(x^0_\mu/d\right)\left(x^0_\nu/d\right)+O\left(\left(\rho_0/d\right)^3\right)\; . \label{Fiuexp}
\ee
In the inner cylinder coordinates ($a<\rho^{\;'}<b $):
\bea
\label{explFiu0}
\Fiu_0(\vec{r}\;{'})=\Vm\frac{\rho^{\;'}-a}{d}\;;\qquad\qquad\qquad\qquad\qquad\qquad\qquad\qquad\qquad\qquad\qquad\;\;\\
\label{explFiumu}
\Fiu_\mu(\vec{r}\;{'})=-\frac{a}{d}\,\Vm \Re\left(c_\mu^+e^{-\imath\vfi^{\;'}}\right)\left(\rho^{\;'}/a-a/\rho^{\;'}\right)
\; ;\qquad\qquad\qquad\qquad\qquad\;\;\;\;\;\;\\
\label{explFiumunu}
\Fiu_{\mu\nu}(\vec{r}\;{'})=\frac{a}{2d}\,\Vm\left\{
\Re\left(c_\mu^+c_\nu^+e^{-2\imath\vfi^{\;'}}\right)\left[\left(\rho^{\;'}/a\right)^2-\left(a/\rho^{\;'}\right)^2\right]+
\delta_{\mu\nu}\frac{\rho^{\;'}-a}{a}
\right\}\; .\qquad\;
\eea

In the outer cylinder coordinates ($a<\rho<b $):
\bea
\label{uexplFiu0}
\Fiu_0(\vec{r})=\Vm\frac{\rho-a}{d}\;; \qquad\qquad\qquad\qquad\qquad\qquad\qquad\qquad\qquad\qquad\qquad\;\;\;\;\;\\
\label{uexplFiumu}
\Fiu_\mu(\vec{r})=-\frac{a}{d}\,\Vm
\Re\left(c_\mu^+e^{-\imath\vfi}\right)
\left(\rho/a-a/\rho\right)
\; ;\qquad\qquad\qquad\qquad\qquad\qquad\quad\;\;\;\\
\label{uexplFiumunu}
\Fiu_{\mu\nu}(\vec{r})=\frac{a}{2d}\,\Vm
\left\{
\Re\left(c_\mu^+c_\nu^+e^{-2\imath\vfi}\right)
\left[\left(\rho/a\right)^2-\left(a/\rho\right)^2\right]+
\delta_{\mu\nu}\frac{\rho-a}{a}
\right\}\; .\qquad\qquad\;\;\;
\eea

Note that $\Fiu_0(\vec{r}\;^{'})$ and $\Fiu_0(\vec{r})$ are identical functions of just different arguments, which, of course, coincide to the lowest order ($\vec{\rho}\;^{'}\equiv\vec{\rho}$ for ${\rho}_0=0$). More importantly, the two--dimensional BVP (\ref{Laplu}) allows for a very simple exact solution in the bipolar coordinates $\{\alpha,\;\beta\}$ (see e.g.~\cite{LSU}, 7.3): it is a linear function of just one coordinate $\alpha$, a transverse coordinate in the asymmetric gap. The symmetric limit (coaxial cylinders) is, however, a singular one for these coordinates, since the focal distance of the coordinate system tends to infinity. Nevertheless, it is still possible to expand the solution properly and reproduce exactly the above results, --- another important check of our calculations.

\section{Electrostatic Energy\label{s3}}

Denote ${\cal D}_L$ a finite domain between the two cylinders cut at $z=\pm L$; ${\cal D}_\infty$ is then the whole infinite volume of the gap between the cylinders. The electrostatic energy stored in ${\cal D}_L$ is expressed through the potential as
\bea
W_{L}=\frac{\epsilon_0}{2}\,\int_{{\cal D}_L}\,\left(\nabla \Phi\right)^2\,dV=
\frac{\epsilon_0}{2}\,\int_{{\cal D}_L}\,dV\,\left[
\left(\nabla \Fiu\right)^2 + 2 \left(\nabla \Fiu\cdot \nabla \Fip \right)+\left(\nabla \Fip\right)^2
\right]\; ,\nonumber
\eea
according to representation (\ref{Fiup}). The energy consists, naturally, of three terms: the one due to the uniform potential, the contribution coming from the interaction between the uniform and patch potentials, and the energy of patches. The uniform potential does not depend on the axial coordinate, so the uniform part of energy (and eventually the force from it, see section \ref{s1.2}) is proportional to the cylinder height, $2L$; i.e., it is the energy per unit length that is finite. We thus write this part of the energy in the form:
\be
W^u(L)=W^u(L,\vec r^{\;0})=\epsilon_0L\,\int_{{\cal C}}\,\left(\nabla_\perp \Fiu\right)^2\,dA\; ; 
\label{WLu}
\ee
${\cal C}$ is the gap annulus at $z=const$, $\nabla_\perp$ is the corresponding two--dimensional gradient.

In the two remaining contributions one can take the limit $L\to\infty$, assuming, as usual, a local character of the patches:
\bea
W^{int}=W^{int}(\vec r^{\;0})=\epsilon_0\,\int_{{\cal D}_\infty}\,\left(\nabla \Fiu\cdot \nabla \Fip \right)\,dV\; ;\label{Wint}\\
W^{p}=W^{p}(\vec r^{\;0})=\frac{\epsilon_0}{2}\,\int_{{\cal D}_\infty}\left(\nabla \Fip\right)^2\,dV\; . \label{Wp}
\eea 
Using the results of the previous section, we are going to calculate the quantities (\ref{WLu})---(\ref{Wp}) one by one in the three sections below; some details of these calculations can be found in Appendix C. In these calculations we systematically employ the following convenient formulas.

Let ${\cal D}$ be some domain with the boundary ${\cal B}$, and $u(\vec r)$ and $v(\vec r)$ be some functions squarely integrable over ${\cal D}$ with their derivatives. Let also $u(\vec r)$ satisfy the Laplace equation in ${\cal D}$; then by integrating by parts we find
\[
0=\int_{{\cal D}}\,v\,\Delta u\,dV=\int_{{\cal B}}\,v\,\frac{\partial u}{\partial n}\,dA-
\int_{{\cal D}}\,\left(\nabla v\cdot \nabla u \right)\,dV\; ,
\]
where $n$ is the outward normal to the boundary ${\cal B}$. Thus
\be
\int_{{\cal D}}\,\left(\nabla v\cdot \nabla u \right)\,dV=\int_{{\cal B}}\,v\,\frac{\partial u}{\partial n}\,dA,\qquad
\Delta u=0\;\;\mbox{in}\;\;{\cal D}\; ; 
\label{uv}
\ee
for $v=u$, in particular,
\be
\int_{{\cal D}}\,\left(\nabla u \right)^2\,dV=\int_{{\cal B}}\,u\,\frac{\partial u}{\partial n}\,dA,\qquad
\Delta u=0\;\;\mbox{in}\;\;{\cal D}\; .
\label{uu}
\ee
Moreover, if both functions are harmonic in ${\cal D}$, then we can represent the integral (\ref{uv}) in two symmetric forms,
\be
\int_{{\cal D}}\,\left(\nabla v\cdot \nabla u \right)\,dV=\int_{{\cal B}}\,v\,\frac{\partial u}{\partial n}\,dA=
\int_{{\cal B}}\,u\,\frac{\partial v}{\partial n}\,dA,\quad
\Delta u=\Delta v=0\;\;\mbox{in}\;\;{\cal D}\; . 
\label{uvsymm}
\ee
Formulas (\ref{uv})---(\ref{uvsymm}) hold in the space of any dimension $D$, in particular, for $D=2,\;3$.

\subsection{Uniform Energy\label{s3.1}}

The definition (\ref{WLu}) and formula (\ref{uu}) with $u=\Fiu$ give:
\[
W^u(L,\vec r^{\;0})=\epsilon_0 L\,\int_{{\cal C}}\left(\nabla_\perp \Fiu\right)^2\,dA=\epsilon_0L
\int^{2\pi}_0bd\vfi\Fiu\frac{\partial \Fiu}{\partial \rho}\Biggl|_{\rho=b}=
\]
\[
\epsilon_0La\Vm
\int^{2\pi}_0d\vfi\frac{\partial \Fiu}{\partial \rho}\Biggl|_{\rho=b}\left[1+O(d/a)\right]\; ,
\]
where we used both boundary conditions (\ref{Laplu}).  Substituting the expansion (\ref{Fiuexp}) for the potential and then carrying out the straightforward integration using expressions (\ref{uexplFiu0})--(\ref{uexplFiumunu}), we obtain ($A_L$ is the area of the piece of cylinder with the height $2L$):
\be
W^u(L,\vec r^{\;0})=W_0^u(L)+W_\mu^u(L)\,(x^0_\mu/d)+\;W_{\mu\nu}^u(L)\,(x^0_\mu/d)\,(x^0_\nu/d)+O\left[(\rho_0/d)^3\right]\; ;
\label{uenexp}
\ee
\bea
W^u_0(L)=\epsilon_0La\Vm\int^{2\pi}_0d\vfi\frac{\partial \Fiu_0}{\partial \rho}\Biggl|_{\rho=b}=
2\pi L \epsilon_0 \frac{a}{d}\left(\Vm\right)^2=\epsilon_0 \frac{A_L\left(\Vm\right)^2}{2d}\; ;\nonumber\\
W^u_\mu(L)=\epsilon_0La\Vm\int^{2\pi}_0d\vfi\frac{\partial \Fiu_\mu}{\partial \rho}\Biggl|_{\rho=b}=0\; ;\label{Wures}\\
W^u_{\mu\nu}(L)=\epsilon_0 La\Vm\int^{2\pi}_0d\vfi\frac{\partial \Fiu_{\mu\nu}}{\partial \rho}\Biggl|_{\rho=b}=
\pi L \epsilon_0 \frac{a}{d}\left(\Vm\right)^2\delta_{\mu\nu}=\epsilon_0 \frac{A_L\left(\Vm\right)^2}{4d}\,\delta_{\mu\nu}\;.\nonumber
\eea
The zero order term here is, of course, the classical energy of a plane capacitor, the same as for the cylindrical one to l.o. in $d/a$; the first order term vanishes due to symmetry.

\subsection{Interaction Energy\label{s3.2}}

We apply formula (\ref{uv}) to the expression (\ref{Wint}) with $u=\Fip$ and $v=\Fiu$, taking into account both boundary conditions in the problem (\ref{Laplu}), which gives:
\[
W^{int}(\vec r^{\;0})=\epsilon_0\,\int_{{\cal D}_\infty}\,\left(\nabla \Fiu\cdot \nabla \Fip \right)\,dV=\epsilon_0a\Vm
\int_{-\infty}^{\infty}dz\int^{2\pi}_0d\vfi\,\frac{\partial \Fip}{\partial \rho}\Biggl|_{\rho=b}\left[1+O(d/a)\right]\;.
\]
As in the uniform case above, we substitute here the expansion (\ref{Fipexp}), to get:
\bea
W^{int}(\vec r^{\;0})=W_0^{int}+W_\mu^{int}\,(x^0_\mu/d)+\;W_{\mu\nu}^{int}\,(x^0_\mu/d)\,(x^0_\nu/d)+O\left[(\rho_0/d)^3\right]\; ;\qquad
\label{intenexp}\\
W^{int}_0=\epsilon_0a\Vm\int_{-\infty}^{\infty}dz\int^{2\pi}_0d\vfi\frac{\partial \Fip_0}{\partial \rho}\Biggl|_{\rho=b};\qquad
W^{int}_\mu=\epsilon_0a\Vm\int_{-\infty}^{\infty}dz\int^{2\pi}_0d\vfi\frac{\partial \Fip_\mu}{\partial \rho}\Biggl|_{\rho=b}\; ;\nonumber\\
W^{int}_{\mu\nu}=\epsilon_0a\Vm\int_{-\infty}^{\infty}dz\int^{2\pi}_0d\vfi\frac{\partial \Fip_{\mu\nu}}{\partial \rho}\Biggl|_{\rho=b}\; .\qquad\qquad\qquad\qquad\qquad
\label{Wintterms}
\eea
The derivatives of the potential at $\rho=b$ are found in Appendix C, formulas (\ref{uDFip0})---(\ref{uDFipmunu}). Each of them has the form of the double Fourier transform (\ref{Four}), 
\[
u(\vfi,z)=\LL\,u_n(k)e^{i(kz+n\vfi)}\; ,
\]
where $u(\vfi,z)$ is any one of these derivatives. By inverting this formula and setting then $k=0$ and $n=0$ in the result, one obtains
\[
u_n(k)=\FF\,u(\vfi,z)e^{-i(kz+n\vfi)},\qquad 
\int_{-\infty}^{\infty}dz\int^{2\pi}_0d\vfi\,u(\vfi,z)=2\pi u_0(0)\; .
\]
The last one is exactly the integral that stands in each of the expressions (\ref{Wintterms}). Thus combining it with the  mentioned formulas for the derivatives, after some not very tedious algebra using the relation $(c_\mu^-)^*=c_\mu^+$ for the coefficients involved, we find:
\bea
W^{int}_0=-2\pi\epsilon_0\frac{a}{d}\Vm\left[
\left(G_0(0)-H_0(0)\right)
\right]\;,\nonumber\qquad\qquad\qquad\qquad\qquad\qquad\qquad\qquad\quad\,\\
\label{Wintres}
W^{int}_\mu=+4\pi\epsilon_0\frac{a}{d}\Vm\Re\left[c_\mu^+\left(G_{1}(0)-H_{1}(0)\right)
\right]\;,\qquad\qquad\qquad\qquad\qquad\qquad\qquad\;\;\;\;\\
W^{int}_{\mu\nu}=-4\pi\epsilon_0\frac{a}{d}\Vm\left\{\Re\left[c_\mu^+ c_\nu^+\left(G_{2}(0)-H_{2}(0)\right)\right]+(\delta_{\mu\nu}/4)\left(G_0(0)-H_0(0)\right)
\right\}\;.\quad\;\,\nonumber
\eea 
In these calculations we have also used the property (\ref{symmGnHn}) as applied to the Fourier coefficient
\be
f_n(k)=G_n(k)e^{\imath k z^0}-H_n(k)\; ;
\label{fn}
\ee 
recall that $\Re(\cdot)$ denotes the real part of $(\cdot)$.

\subsection{Patch Energy\label{s3.4}}

We now apply the formula (\ref{uu}) with $u=\Fip$ to the definition (\ref{Wp}) of the patch energy. Making use of the boundary conditions (\ref{bcap}) and (\ref{bcbp}), we obtain:
\[
W^{p}(\vec r^{\;0})=
\frac{\epsilon_0}{2}\int_{{\cal D}_\infty}\,
\left(\nabla \Fip\cdot \nabla \Fip \right)\,dV=
\]
\[
\frac{\epsilon_0}{2}\left\{
\int_{-\infty}^{\infty}dz\int^{2\pi}_0 bd\vfi\,H(\vfi,z)\frac{\partial \Fip}{\partial \rho}\Biggl|_{\rho=b}-
\int_{-\infty}^{\infty}dz\;^{'}\int^{2\pi}_0 a d\vfi\;^{'}\,G(\vfi^{\;'},z^{\;'})\frac{\partial \Fip}{\partial \rho\;^{'}}\Biggl|_{\rho\;^{'}=a}
\right\}\;.\label{Wpt}\quad
\]
Combining this with the expansion (\ref{Fipexp}) for the patch potential we arrive at the energy of patches written as in the formula (\ref{engen}):
\be
W^{p}(\vec r^{\;0})=W_0^{p}+W_\mu^{p}\,(x^0_\mu/d)+\;W_{\mu\nu}^{p}\,(x^0_\mu/d)\,(x^0_\nu/d)+O\left[(\rho_0/d)^3\right]\; ,
\label{penexp}
\ee
where, to l. o. in $d/a$,
\be
W^{p}_\xi=\frac{\epsilon_0 a}{2}\left\{
\int_{-\infty}^{\infty}dz\int^{2\pi}_0 d\vfi\,H(\vfi,z)\frac{\partial \Fip_\xi}{\partial \rho}\Biggl|_{\rho=b}-
\int_{-\infty}^{\infty}dz\;^{'}\int^{2\pi}_0 d\vfi\;^{'}\,G(\vfi^{\;'},z^{\;'})\frac{\partial \Fip_0}{\partial \rho\;^{'}}\Biggl|_{\rho\;^{'}=a}
\right\}\;,\quad\;\;\label{Wpterms}
\ee
for $\xi=0,\;\mu,\;\mu\nu$, i. e., for all the functional coefficients in the expression (\ref{penexp}). These double integrals can be calculated immediately employing the Parceval identity (\ref{Parc}), for which one needs the Fourier coefficients of each of the two factors in the integrand in (\ref{Wpterms}). The first factors $H(\vfi,z)$, $G(\vfi^{\;'},z^{\;'})$ give no such problem, while the second ones are the derivatives of the potentials at the boundaries. Their expressions are found in Appendix C: formulas (\ref{uDFip0}), (\ref{uDFipmunu}) provide them at $\rho=b$, while (\ref{DFip0}), (\ref{DFipmunu}) work for $\rho{\;'}=a$. All these expressions have a structure of the double Fourier transform with the explicit Fourier coefficients. Therefore we can write, first for the zeroth order term in the energy expansion (\ref{penexp}):
\[
W^{p}_0=\frac{\epsilon_0 a}{2}\int_{-\infty}^{\infty}dk\sum_{n=-\infty}^{\infty}\left\{
H_n^{*}(k)\left[
-\frac{f_n(k)}{d}\right]-G_n^{*}(k)
\left[-\frac{f_n(k)}{d}\right]e^{-\imath k z^0}
\right\}\;=\qquad\qquad\qquad\qquad
\]
\be
\label{W0contr}
\frac{\epsilon_0 a}{2d}\int_{-\infty}^{\infty}dk\sum_{n=-\infty}^{\infty}|{f_n(k)}|^2\; ,
\ee
where we used the notation (\ref{fn}) again, as well as the formulas (\ref{uDFip0}) and (\ref{DFip0}).

Similar calculations of $W^{p}_\mu,\;W^p_{\mu\nu}$ are based on the  formulas (\ref{uDFip0}), (\ref{DFip0}), and (\ref{uDFipmunu}), (\ref{DFipmunu}) respectively; they exploit relations $\sum_{n=-\infty}^{\infty}c_\mu^-f_n^{*}(k)f_{n-1}(k)=\sum_{n=-\infty}^{\infty}c_\mu^-f_{n+1}^{*}(k)f_n(k)$ and $(c_\mu^-)^{*}=c_\mu^+$ resulting in
\bea
\label{Wmucontr}
W^{p}_\mu=-\frac{\epsilon_0 a}{2d}\int_{-\infty}^{\infty}dk\sum_{n=-\infty}^{\infty}2\Re
{\left[c_\mu^+f_n^{*}(k)f_{n+1}(k)
\right]}\;;\\
W^{p}_{\mu\nu}=\frac{\epsilon_0 a}{2d}\int_{-\infty}^{\infty}dk\sum_{n=-\infty}^{\infty}
\left[
\delta_{\mu\nu}/2|f_n(k)|^2+2\Re{\left( c_\mu^+c_\nu^+f_n^{*}(k)f_{n+2}(k)\right)}
\right] \;.\nonumber
\eea
Replacing now $f_n(k)$ with its expression (\ref{fn}) in the formulas (\ref{W0contr}), (\ref{Wmucontr}) allows us to obtain the final answer for the energy of the patches:
\bea
\label{Wpres}
W^{p}_0=\frac{\epsilon_0 a}{2d}\int_{-\infty}^{\infty}dk\sum_{n=-\infty}^{\infty}|G_n(k)e^{\imath k z^0}-H_n(k)|^2\;;\nonumber\qquad\qquad\qquad\qquad\qquad\qquad\qquad\qquad\;\;\;\;\,\\
W^{p}_\mu=-\frac{\epsilon_0 a}{d}\int_{-\infty}^{\infty}dk\sum_{n=-\infty}^{\infty}\,\Re\left[c_\mu^+\left(G_n^{*}(k)e^{-\imath k z^0}-H_n^{*}(k)\right)\left(G_{n+1}(k)e^{\imath k z^0}-H_{n+1}(k)\right)\right]\;;\quad\;\,\\
W^{p}_{\mu\nu}=\frac{\epsilon_0 a}{2d}\int_{-\infty}^{\infty}dk\sum_{n=-\infty}^{\infty}\,\left\{
\delta_{\mu\nu}/2|G_n(k)e^{\imath k z^0}-H_n(k)|^2+\right.\nonumber\qquad\qquad\qquad\qquad\qquad\qquad\qquad\\
\left.2\Re\left[c_\mu^+c_\nu^+\left(G_n^{*}(k)e^{-\imath k z^0}-H_n^{*}(k)\right)\left(G_{n+2}(k)e^{\imath k z^0}-H_{n+2}(k)\right)
\right]
\right\}\nonumber\;.\;\;\;
\eea
Remarkably, the expression for the zeroth order energy $W^{p}_0$ is perfectly similar to that in the spherical case [see~\cite{KKS}, the last expression in the formula (3) there]. The similarity gets even more pronounced when rewriting the formula (\ref{Wpres}) for $W^{p}_0$ using the Parceval identity: \[
W^{p}_0=\frac{\epsilon_0 a}{2d}\int_{-\infty}^{\infty}dz \int_{0}^{2\pi}d\vfi\left[V_a(\vfi,z)-V_b(\vfi,z)\right]^2\; ;
\]
here $V_a(\vfi,z)\equiv G(\vfi,z)$ and $V_b(\vfi,z)\equiv H(\vfi,z)$ are the patch voltages taken in the coordinate of the outer cylinder. 

The calculation of the energy is now finished; we have all its parts in exactly the form needed for the calculation of the electrostatic force.

\begin{acknowledgments}
This work was supported by ICRANet (V.F.) and by KACST through the collaborative agreement with GP-B (A.S.). The authors are grateful to Remo~Ruffini and Francis~Everitt for their permanent interest in and support of this work, and for some valuable remarks. John~Mester has helped us a lot by providing information about STEP, and by encouraging to carry out this analysis. A special thanks to David Hipkins, a great motivator for the patch effect analysis. We are extremely greatful to Paul Worden for his input regarding the STEP requirements, and other remarks. We thank Dan DeBra and Sasha Buchman for fruitfull discussions and insightful suggestions.
\end{acknowledgments}

\appendix
\section{Calculation of the Patch Potential}\label{A1}

The re--expansion formulas for cylindrical solutions to the Laplace equation in shifted coordinates are found in~\cite{Ye}, {\bf 131.2.2} and {\bf 136.2.2}. They read:
\bea
I_n(k\rho^{\;'})\,e^{in\vfi^{\;'}}=\sum\limits_{m=-\infty}^{\infty}\,I_{n-m}(k\rho_0)e^{i(n-m)\vfi_0}\,I_m(k\rho)\,e^{im\vfi}\; ;\nonumber\\
K_n(k\rho^{\;'})\,e^{in\vfi^{\;'}}=\sum\limits_{m=-\infty}^{\infty}\,(-1)^{n-m}I_{n-m}(k\rho_0)e^{i(n-m)\vfi_0}\,K_m(k\rho)\,e^{im\vfi}\; ,\nonumber
\eea
where $\rho_0$ and $\vfi_0$ are the polar components of the shift as defined in (\ref{engen}) and (\ref{cyl_shift}). Therefore
\bea
\sum\limits_{n=-\infty}^{\infty}\left[A_n(k)\,I_n(k\rho^{\;'})+B_n(k)\,K_n(k\rho^{\;'})\right]e^{in\vfi^{\;'}}=\nonumber\\
\sum\limits_{n=-\infty}^{\infty}\,I_n(k\rho)\,e^{in\vfi}\sum\limits_{m=-\infty}^{\infty}\,I_{m-n}(k\rho_0)e^{i(m-n)\vfi_0}A_m(k)+\nonumber\\
\sum\limits_{n=-\infty}^{\infty}\,K_n(k\rho)\,e^{in\vfi}\sum\limits_{m=-\infty}^{\infty}\,(-1)^{m-n}I_{m-n}(k\rho_0)e^{i(m-n)\vfi_0}B_m(k)\; ,\nonumber
\eea
where we have changed the order of summation and then switched the indeces $m$ and $n$. When substituted in the representation (\ref{preprpr}) of the patch potential $\Fip$ in primed coordinates taking into account $z^{'}=z+z^0$, this gives exactly the representation (\ref{preprunpr})---(\ref{tildBn}) in the unprimed coordinates of the outer cylinder.

We are now to construct the 2nd order perturbative solution (\ref{AnBn}),
\be
\left[
\matrix{\,A_n(k)\cr    
				\,B_n(k)\cr}
\right]=
\left[
\matrix{\,A^0_n(k)\cr
        \,B^0_n(k)\cr}
\right]+
\left[
\matrix{\,A^{\mu}_n(k)\cr
        \,B^{\mu}_n(k)\cr}
\right]\,x^0_\mu+
\left[
\matrix{\,A^{\mu\nu}_n(k)\cr
        \,B^{\mu\nu}_n(k)\cr}
\right]x^0_\mu\,x^0_\nu+
O\left(\rho_0^3\right)\; ,
\label{solAnBn}
\ee
to the infinite system of linear algebraic equations (\ref{infsys}). As mentioned in section \ref{s2.2}, its matrix coefficients contain the factor $I_{m-n}(k\rho_0)\sim O\left(\rho_0^{|m-n|}\right)$, so only the terms $m=~n, \; n\pm~1,\; n\pm~2$ should be taken into account, to the 2nd order. Using thus the formulas (all the needed results on Bessel functions can be found in ~\cite{Leb}, Ch. 5)
\bea
I_{0}(k\rho_0)=1+(k\rho_0)^2/4+O\left((k\rho_0)^4\right),\qquad I_{\pm\,1}(k\rho_0)=(k\rho_0)/2+O\left((k\rho_0)^3\right)\; ,\nonumber\\
I_{\pm\,2}(k\rho_0)=(k\rho_0)^2/8+O\left((k\rho_0)^4\right)\; ,\qquad\qquad\qquad\qquad\nonumber
\eea
as well as the convenient representations ($c_1^+\equiv0.5,\quad c_2^+\equiv0.5i$)
\be
\rho_0^2=\delta_{\mu\nu}x^0_\mu\,x^0_\nu;\qquad (\rho_0/2)e^{i\vfi_0}=c_\mu^+x^0_\mu,\qquad 
(\rho_0/2)^2e^{2i\vfi_0}=c_\mu^+c_\nu^+x^0_\mu\,x^0_\nu\; ,\label{c12}
\ee
we find:
\be
I_{0}(k\rho_0)=1+\frac{k^2}{4}\delta_{\mu\nu}x^0_\mu\,x^0_\nu+O\left((k\rho_0)^4\right)\label{asympI}\; ;
\ee
\bea
I_{1}(k\rho_0)e^{i\vfi_0}=k\,c_\mu^+x^0_\mu+O\left((k\rho_0)^3\right),\quad
I_{-1}(k\rho_0)e^{-i\vfi_0}=k\,c_\mu^- x^0_\mu+O\left((k\rho_0)^3\right),\quad c_\mu^-\equiv(c_\mu^+)^*\nonumber\; ;\;\;\qquad\qquad\qquad\qquad\\
I_{2}(k\rho_0)e^{2i\vfi_0}=\frac{k^2}{2}c_\mu^+c_\nu^+x^0_\mu\,x^0_\nu+O\left((k\rho_0)^4\right)\,,\qquad 
I_{-2}(k\rho_0)e^{-2i\vfi_0}=\frac{k^2}{2}c_\mu^- c_\nu^-x^0_\mu\,x^0_\nu+O\left((k\rho_0)^4\right)\nonumber\; .\qquad\qquad\qquad\qquad\;\; 
\eea
Here we introduced the notation $c_\mu^-$ for a uniform writing. Substituting expressions (\ref{solAnBn}) and (\ref{asympI}) in the system (\ref{infsys}) allows us to obtain the following sequence of equations for the unknown coefficients in expansions (\ref{solAnBn}): 
\[
\cases{I_n(ka)\,A_n^\xi(k)\,+K_n(ka)\,B_n^\xi(k)\;=\;G_n^\xi(k)\; ,\cr
       I_n(kb)\,A_n^\xi(k)\,+K_n(kb)\,B_n^\xi(k)\;=\;H_n^\xi(k)\; ,\cr}
\]
where $n=0,\pm1,\pm2,\ldots$, and the index $\xi$ assumes three values $\xi=0,\;\mu,\;\mu\nu$, with  $\mu,\nu=1,2$.
The right hand sides here are:
\bea
\label{Hnmu}
G_n^0(k)=G_n(k),\qquad \qquad G_n^{\mu}(k)=G_n^{\mu\nu}=0\; ;\label{Hnmunu}\qquad\qquad\qquad\qquad\\
H_n^0(k)=H_n(k)\,e^{-ikz^0},\qquad H_n^\mu(k)=-k\,c_\mu^\pm\left[I_n(kb)\,\,A_{n\pm\,1}^0(k)-K_n(kb)\,\,B_{n\pm\,1}^0(k)\right]\; ;\nonumber\\
H_n^{\mu\nu}(k)=-k\left[I_n(kb)\,c_\mu^\pm\,A_{n\pm\,1}^\nu(k)-K_n(kb)\,c_\mu^\pm\,B_{n\pm\,1}^\nu(k)\right]-\nonumber\qquad\qquad\qquad\\
\frac{k^2}{2}\left[I_n(kb)\left(c_\mu^\pm\,c_\nu^\pm\,A_{n\pm\,1}^0(k)+\frac{\delta_{\mu\,\nu}}{2}\,A_n^0(k)\right)+
K_n(kb)\left(c_\mu^\pm\,c_\nu^\pm\,B_{n\pm\,1}^0(k)+\frac{\delta_{\mu\,\nu}}{2}\,B_n^0(k)\right) \right] 
\,,\nonumber
\eea
and terms like $c^\pm\,A_{n\pm\,1}$ and $c^\pm c^\pm\,A_{n\pm\,1}$ should be read as
\[
c^\pm\,A_{n\pm\,1}(k)=c^+\,A_{n+1}(k)+c^-\,A_{n-1}(k)\,, \qquad c^\pm c^\pm\,A_{n\pm\,1}(k)=c^+c^+\,A_{n+1}(k)+c^-c^-\,A_{n-1}(k)\,.
\]\,
Solving the above linear systems leads to the following set of answers ($\xi=\mu,\;\mu\nu$):
\bea
A_n^{0}(k)=\frac{G_n(k)K_n(kb)-H_n(k)K_n(ka)e^{-\imath\,kz^0}}{D_n(k)}\;,\nonumber\\ 
B_n^{0}(k)=\frac{H_n(k)I_n(ka)e^{-\imath\,kz^0}-G_n(k)I_n(kb)}{D_n(k)}\; ;\label{AB0}\\
A_n^{\xi}(k)=-\frac{H_n^\xi(k)K_n(ka)}{D_n(k)},\;\qquad 
B_n^{\xi}(k)=\frac{H_n^\xi(k)I_n(ka)}{D_n(k)}\; ;\label{ABxi}
\eea
where 
\be
D_n(k)=K_n(kb)\,I_n(ka)-K_n(ka)\,I_n(kb)\; .
\label{det}
\ee

Expressions (\ref{AB0})---(\ref{det}) and (\ref{solAnBn}) provide the solution for the patch potential in the form (\ref{preprpr}) to the 2nd order in $\rho_0/d$ and {\it any} value of $d/a$. We now simplify all the expressions to l. o. in this small second parameter. To do this, we employ the Taylor expansions:
\bea
I_n(kb)&=&I_n(ka)+ I_n^{\;'}(ka)\,kd+O\left[\left(kd\right)^2\right],\qquad b=a+d,\quad d \to\,0\; ;\nonumber\\
K_n(kb)&=&K_n(ka)+ K_n^{\;'}(ka)\,kd+O\left[\left(kd\right)^2\right]\,;\label{kbtoka}
\eea
where the primes denote the derivatives with respect to the whole argument. With the help of these expansions and the known formula for the Wronskian of the Bessel functions,
\be
\label{Wronskian}
W\left(I_n(\xi),K_n(\xi)\right)=-1/\xi\,,
\ee
we simplify the denominator (\ref{det}) to lowest order in $d/a$ as:
\bea
D_n(k)=kd\left[I_n(ka)K_n^{\,'}(ka)-K_n(ka)I_n^{\,'}(ka)\right]+\frac{d}{a}\left[O\left(\frac{d}{a}\right)+O\left(\left(n(ka)\frac{d}{a}\right)^2\right)
\right]=\nonumber\\
-\frac{d}{a}\left[1+O\left(\frac{d}{a}\right)+O\left(\left(n(ka)\frac{d}{a}\right)^2\right)\right]\;.\quad\label{Adet}
\eea
The second term in the remainder could be dropped if $k$ and $n$ were bounded. But in our formulas both run from minus to plus infinity, so we need to keep this remainder to make the asymptotic expressions uniform for the whole range of $k$ and $n$. However, when substituted in the sums over $n$ and integrals over $k$, as in formula (\ref{preprpr}), these terms give rise to contributions $O((d/a)^2)$ provided that the proper sums/integrals converge,
\be
\label{L2d}
\int_{-\infty}^{\infty}dk\sum_{n=-\infty}^{\infty} k^2\,n^2 |G_n(k)|^2<\infty \,,\qquad \int_{-\infty}^{\infty}dk\sum_{n=-\infty}^{\infty}  k^2\,n^2 |H_n(k)|^2<\infty\; .
\ee
By the Parceval identity (\ref{Parc}), the latter conditions are equivalent to
\be
\int\limits_{0}^{2\pi}\int\limits_{-\infty}^{\infty}d\vfi^{\;'}dz^{\;'}\,|\partial^2 G/\partial\vfi^{\;'}\partial z^{\;'})|^2<\infty,\qquad
\int\limits_{0}^{2\pi}\int\limits_{-\infty}^{\infty}d\vfi dz\,|\partial^2 H/\partial\vfi\partial z)|^2<\infty\; .
\label{L2der}
\ee
Moreover, in the expressions of $A_n^0,\;A_n^\mu,\;A_n^{\mu\nu}$ we use the expansion
\[
K_n(ka)=K_n(kb)- K_n^{\;'}(ka)\,kd+O\left[\left(kd\right)^2\right]\, ,
\]
while for $B_n^0$ we employ the first of the formulas (\ref{kbtoka}). This is done, in fact, to provide the proper convergence of integrals over $k$ and series in $n$ at infinity, see section \ref{s2.2}.

For the zero order coefficients, we thus have:
\bea
A_n^{0}(k)=-\frac{a}{d}\left\{
K_n(kb)\left[G_n(k)-H_n(k)e^{-\imath\,kz^0}\right]+O\left(\frac{d}{a}\right)+O\left(\left(n(ka)\frac{d}{a}\right)^2\right)
\right\}\; ,\nonumber\\
B_n^{0}(k)=\frac{a}{d}\left\{
I_n(ka)\left[G_n(k)-H_n(k)e^{-\imath\,kz^0}\right]+O\left(\frac{d}{a}\right)+O\left(\left(n(ka)\frac{d}{a}\right)^2\right)
\right\}\; .\qquad\label{AAB0}
\eea
From now on, all our asymptotic formulas have the same two--term remainder estimates, which we simply drop.

By the formulas (\ref{AAB0}) and definition (\ref{Hnmu}), we obtain thus
\[
H_n^\mu(k)=\frac{1}{d}
c_\mu^\pm\left[G_{n\pm\,1}(k)-H_{n\pm\,1}(k)e^{-\imath\,kz^0}\right]
\; ,
\]
and the expressions (\ref{ABxi}) become
\bea
A_n^{\mu}(k)=\,\frac{a}{d^{\,2}}
K_n(kb)\,c_\mu^\pm\left[G_{n\pm\,1}(k)-H_{n\pm\,1}(k)e^{-\imath\,kz^0}\right]
\; ,\nonumber\\
B_n^{\mu}(k)=\,-\frac{a}{d^{\,2}}
I_n(ka)\,c_\mu^\pm\left[G_{n\pm\,1}(k)-H_{n\pm\,1}(k)e^{-\imath\,kz^0}\right]
\,.\label{AABmu}
\eea
By the same token, formula (\ref{Hnmunu}) simplifies to
\bea
H_n^{\mu\nu}(k)=\frac{1}{d^2}\left\{
c_\mu^\pm\,c_\nu^\pm\left[G_{n\pm\,2}(k)-H_{n\pm\,2}(k)e^{-\imath\,kz^0}\right]+\frac{\delta_{\mu\nu}}{2}\left[G_n(k)-H_n(k)e^{\imath\,kz^0}\right]
\right\}\;,\qquad\nonumber
\eea
and the coefficients (\ref{ABxi}) are written as
\bea
A_n^{\mu\nu}(k)=\qquad\qquad\qquad\qquad\qquad\qquad\qquad\qquad\qquad\qquad\qquad\qquad\qquad\qquad\qquad\qquad\qquad\qquad\nonumber\\
-\frac{a}{d^{\,3}}\left\{
K_n(kb)\left[c_\mu^\pm\,c_\mu^\pm\left[G_{n\pm\,2}(k)-H_{n\pm\,2}(k)e^{-\imath\,kz^0}\right]+\frac{\delta_{\mu\,\nu}}{2}\left[G_n(k)-H_n(k)e^{-\imath\,kz^0}\right]\right]
\right\}\;;\label{AABmunu}\qquad\\
B_n^{\mu\nu}(k)=\nonumber\qquad\qquad\qquad\qquad\qquad\qquad\qquad\qquad\qquad\qquad\qquad\qquad\qquad\qquad\qquad\qquad\qquad\qquad\\
\frac{a}{d^{\,3}}\left\{
I_n(ka)\left[c_\mu^\pm\,c_\mu^\pm\left[G_{n\pm\,2}(k)-H_{n\pm\,2}(k)e^{-\imath\,kz^0}\right]+\frac{\delta_{\mu\,\nu}}{2}\left[G_n(k)-H_n(k)e^{-\imath\,kz^0}\right]\right]
\right\}\;.\nonumber\qquad
\eea
Expressions (\ref{AAB0})---(\ref{AABmunu}), when substituted in the representation (\ref{preprpr}), provide exactly the formulas (\ref{explFip0})--(\ref{explFipmunu}) for the potential in the primed coordinates. 

To obtain the potential in the unprimed coordinates we need, as seen from the formula (\ref{preprunpr}), to compute the coefficients $\tilde{A_n},\tilde{B_n}$ to the 2nd order in the transverse shift, in a complete similarity with the formula (\ref{solAnBn}):
\be
\left[
\matrix{\,\tilde{A}_n(k)\cr    
				\,\tilde{B}_n(k)\cr}
\right]=
\left[
\matrix{\,\tilde{A}^0_n(k)\cr
        \,\tilde{B}^0_n(k)\cr}
\right]+
\left[
\matrix{\,\tilde{A}^{\mu}_n(k)\cr
        \,\tilde{B}^{\mu}_n(k)\cr}
\right]\,x^0_\mu+
\left[
\matrix{\,\tilde{A}^{\mu\nu}_n(k)\cr
        \,\tilde{B}^{\mu\nu}_n(k)\cr}
\right]x^0_\mu\,x^0_\nu+
O\left(\rho_0^3\right)\; ,
\label{soltAnBn}
\ee
To get them, we simply use the definitions (\ref{tildAn}) and (\ref{tildBn}) of the coefficients $\tilde{A_n},\tilde{B_n}$, and introduce there the expansions (\ref{solAnBn}) with the known coefficients (\ref{AAB0})---(\ref{AABmunu}), as well as the expansions (\ref{asympI}). This is a tedious and cumbersome, but straightforward calculation, which ends with the representations (\ref{uexplFip0})--(\ref{uexplFipmunu}).

\section{Calculation of the Potential due to\\ the Uniform Boundary Voltages} \label{A2}

The uniform potential is obtained substituting the Fourier coefficients of the boundary functions (\ref{gnhnu}),
\[
G_n(k)=0\;,\qquad H_n(k)=2\pi \Vm\delta(k)\delta_{n0}\; ,
\]
in the formulas found for the patch potential, (\ref{explFip0})--(\ref{explFipmunu}) and (\ref{uexplFip0})--(\ref{uexplFipmunu}). 
The zeroth order (coaxial configuration) potential in the primed coordinates then is [recall the definitions (\ref{Omega}) and (\ref{NotFour})]:
\[
\Fiu_0(\vec{r}\;^{'})=
\frac{a}{d}\,\Vm\lim_{k\to 0}\left[
K_0(kb)I_0(k\rho^{\;'})-I_0(ka)K_0(k\rho^{\;'})
\right]\;.
\]
Since $I_0(\xi)\to1,\;K_0(\xi)\sim\ln(2/\xi)$ when $\xi\to0$, we furthermore obtain
\be
\Fiu_0(\vec{r}^{\;'})=\frac{a}{d}\,\Vm\,\ln (\rho\;^{'}/b)=
\frac{a}{d}\,\Vm\,\left[\ln (\rho\;^{'}/a)+O(d/a)\right]\; .
\label{Fiulog}
\ee
The logarithm above is a remnant of the exact solution for the coaxial case, $[\Vm\,\ln (\rho\,^{'}/a)/\ln (b/a)]$, with the logarithm in the denominator taken to lowest order in $d/a\ll 1$ (recall that $d=b-a$). Moreover, the upper log may be simplified in a similar way:
\[
\ln(\rho^{\;'}/a)=\ln\left(1+\frac{\rho^{\;'}-a}{a}\right)=\frac{\rho^{\;'}-a}{a}\left[1+O(d/a)\right]\; ;\qquad\qquad\qquad\qquad\qquad\qquad\quad
\]
the potential in a {\it narrow} capacitor does not feel the curvature of the electrodes to l. o., therefore it is a linear function  of the transverse coordinate. Introducing the last expression to the formula (\ref{Fiulog}) we finally obtain:
\be
\Fiu_0(\vec{r}^{\;'})=\Vm\,\frac{\rho^{\;'}-a}{d}\left[1+O(d/a)\right]
\; .
\label{Fiu0fin}
\ee
 
By the same token, the linear and quadratic potentials are obtained from expressions (\ref{explFipmu}) and (\ref{explFipmunu}) as:
\[
\Fiu_\mu(\vec{r}\;^{'})=-\frac{a}{d}\,\Vm\lim_{k\to 0}\left\{
c_\mu^\pm e^{\mp\,\imath\vfi^{\;'}}\left[K_{\mp\,1}(k b)I_{\mp\,1}(k \rho^{\;'})-
I_{\mp\,1}(k a)K_{\mp\,1}(k \rho^{\;'})\right]
\right\}\;,
\]
and
\bea
\Fiu_{\mu\nu}(\vec{r}\;^{'})=\frac{a}{d}\,\Vm\lim_{k\to 0}\left\{
c_\mu^\pm c_\nu^\pm e^{\mp\,2\imath\vfi^{\;'}}\left[K_{\mp\,2}(k b)I_{\mp\,2}(k \rho^{\;'})-
I_{\mp\,2}(k a)K_{\mp\,2}(k \rho^{\;'})\right]+\right.\nonumber\\
\left.\delta_{\mu\nu}/2\left[K_0(k b)I_0(k \rho^{\;'})-I_0(k a)K_0(k \rho^{\;'})\right]
\right\}\;.\nonumber
\eea
Since $I_{-n}(\xi)=I_{n}(\xi),\;K_{-n}(\xi)=K_{n}(\xi)$, these may be written as ($\Re(\cdot)$ is the real part of $(\cdot)$):
\be
\label{AsymFiumu}
\Fiu_\mu(\vec{r}\;^{'})=-\frac{a}{d}\,\Vm\lim_{k\to 0}\left\{2\Re(c_\mu^+ e^{-\imath\vfi^{\;'}})
\left[K_1(kb)I_1(k\rho^{\;'})-
I_1(ka)K_1(k\rho^{\;'})\right]\right\}\; ;
\ee
\bea
\label{AsymFiumunu}
\Fiu_{\mu\nu}(\vec{r}\;^{'})=\frac{a}{d}\,\Vm\lim_{k\to 0}\left\{
2\Re(c_\mu^+c_\nu^+ e^{-2\imath\vfi^{\;'}})\left[K_2(kb)I_2(k\rho^{\;'})-I_2(ka)K_2(k\rho^{\;'})\right]+\right.\nonumber\\
\left.\delta_{\mu\nu}/2\left[K_0(kb)I_0(k\rho^{\;'})-I_0(ka)K_0(k\rho^{\;'})\right]
\right\}\; .
\eea
The near zero asymptotics of Bessel functions allows one to transform (\ref{AsymFiumu}), (\ref{AsymFiumunu}) to:
\bea
\label{Fiumufin}
\Fiu_\mu(\vec{r}\;^{'})=-\frac{a}{d}\,\Vm\left\{\Re(c_\mu^+ e^{-\imath\vfi^{\;'}})
\left[\rho^{\;'}/b-a/\rho^{\;'}\right]\right\}=\nonumber\\
-\frac{a}{d}\,\Vm\left\{\Re(c_\mu^+ e^{-\imath\vfi^{\;'}})
\left[\rho^{\;'}/a-a/\rho^{\;'}\right]+O\left(d/a\right)
\right\}\; ;
\eea
\bea
\label{Fiumunufin}
\Fiu_{\mu\nu}(\vec{r}\;^{'})=\frac{a}{2d}\,\Vm\Re\left\{
c_\mu^+c_\nu^+e^{-2\imath\vfi^{\;'}}\left[\left(\rho^{\;'}/b\right)^2-\left(a/\rho^{\;'}\right)^2\right]+
\delta_{\mu\nu}\ln\left(\rho^{\;'}/b\right)
\right\}=\nonumber\\
\frac{a}{2d}\,\Vm\Re\left\{
c_\mu^+c_\nu^+e^{-2\imath\vfi^{\;'}}\left[\left(\rho^{\;'}/a\right)^2-\left(a/\rho^{\;'}\right)^2\right]+
\delta_{\mu\nu}\frac{\rho^{\;'}-a}{a}+O\left(d/a\right)
\right\}\; ,
\eea
where we used the same transformation of a logarithm as when deriving formula (\ref{Fiu0fin}) from expression (\ref{Fiulog}).

Formulas (\ref{Fiu0fin}), (\ref{Fiumufin}), and (\ref{Fiumunufin}) are the final expressions for the uniform potential in the inner cylinder coordinates as given in formulas (\ref{explFiu0})--(\ref{explFiumunu}). The potential in the unprimed coordinates is now gotten in exactly the same way without any new difficulties: one just starts, respectively, from the expressions (\ref{uexplFip0})--(\ref{uexplFipmunu}) of the patch potential in the unprimed coordinates, and follows the steps described above; in this way, the final results (\ref{uexplFiu0})--(\ref{uexplFiumunu}) are found.

\section{Calculation of the Energy} \label{A3}

Here we provide  the intermediate results needed for the energy computation. The first of them is the derivatives $\partial\Fip_0/\partial \rho,\;\partial\Fip_\mu/\partial \rho,\; \partial\Fip_{\mu\nu}/\partial \rho$ at the outer cylinder boundary $\rho=~b$ required to calculate $W^{int}$ by the formulas (\ref{Wintterms}). First of all we differentiate in $\rho$ the expressions (\ref{uexplFip0})--(\ref{uexplFipmunu}), i.e., the function $\Omega_n(k\rho)$ in them. The result is:
\be
\frac{\partial\Omega_n(k\xi)}{\partial \xi}=\frac{\partial}{\partial \xi}\left[K_n(kb)I_n(k\xi)-I_n(ka)K_n(k\xi)\right]=
 k\left[K_n(kb)I_n^{\;'}(k\xi)-I_n(ka)K_n^{\;'}(k\xi)\right],\label{derOm}
\ee
where the prime indicates the derivative with respect to the whole argument, as usual. 
At the outer boundary, exploiting the relation
\[
I_n(ka)=I_n(kb)-kd I_n^{\;'}(ka)+O\left((kd)^2\right)\; ,
\]
and the formula (\ref{Wronskian}) for the Wronskian, we can write the derivative (\ref{derOm})
\[
\frac{\partial\Omega_n(k\xi)}{\partial \xi}\Biggl|_{\xi=b}=
k\left[-W(I_n(k\xi),K_n(k\xi))|_{k\xi=kb}+O[(kd)^2] \right]=
\frac{1}{b}+O\left[k^3a^2\left(\frac{d}{a}\right)^2\right]\;.
\]
Combining this with the representations (\ref{uexplFip0})---(\ref{uexplFipmunu}), we obtain:
\bea
\frac{\partial\Fip_0}{\partial \rho}\Biggl|_{\rho=b}=-\frac{1}{d}\,{\cal F}
\left\{
f_n(k)\,e^{\imath\left(kz+n\vfi\right)}
\right\};\quad \frac{\partial\Fip_\mu}{\partial \rho}\Biggl|_{\rho=b}=\frac{1}{d}\,{\cal F}
\left\{
c_\mu^\pm f_{n\pm\,1}(k)\,e^{\imath\left(kz+n\vfi\right)}
\right\}\;;\label{uDFip0}\\
\frac{\partial\Fip_{\mu\nu}}{\partial \rho}\Biggl|_{\rho=b}=-\frac{1}{d}\,{\cal F}\left\{
\left[
c_\mu^\pm\,c_\nu^\pm f_{n\pm\,2}(k)+\left(\delta_{\mu\,\nu}/2\right)f_n(k)
\right]e^{\imath\left(kz+n\vfi\right)}\right\}\;.\label{uDFipmunu}\qquad\qquad\quad
\eea
Here ${\cal F}$ is the integration---summation operator defined in (\ref{NotFour}), and, for brevity, 
\be
\label{fndef}
f_n(k)\equiv G_n(k)e^{\imath k z^0}-H_n(k)\; .
\ee
The correction in the formulas (\ref{uDFip0}), (\ref{uDFipmunu}) is given by  $\left[1+O(d/a)\right]$ under the conditions
\be
\int\limits_{-\infty}^{\infty}\,dk\sum\limits_{n=-\infty}^{\infty}\,|k|^3|G_n(k)|<\infty,\qquad 
\int\limits_{-\infty}^{\infty}\,dk\sum\limits_{n=-\infty}^{\infty}\,|k|^3|H_n(k)|<\infty\; . 
\label{newcond}
\ee
The derivatives   $\;\partial\Fip_0/\partial \rho^{\;'},\;\partial\Fip_\mu/\partial \rho^{\;'},\; \partial\Fip_{\mu\nu}/\partial \rho^{\;'}\;$ at the inner cylinder $\rho^{\;'}=a$ are also needed, by the expressions (\ref{Wpterms}). They are derived from the formulas (\ref{explFip0})---(\ref{explFipmunu}) exactly as above, the only significant variation is in transforming the formula (\ref{derOm}). Using the second of expansions (\ref{kbtoka}) and formula (\ref{Wronskian}), we find:
\[
\frac{\partial\Omega_n(k\xi)}{\partial \xi}\Biggl|_{\xi=a}=
 k\left[-W(I_n(k\xi),K_n(k\xi))|_{k\xi=ka}+O[(kd)^2] \right]=
 \frac{1}{a}+O\left[k^3a^2\left(\frac{d}{a}\right)^2\right]\;.
\]
 With this small change, the desired formulas are found as:
\bea
d\,\frac{\partial\Fip_0}{\partial \rho^{\;'}}\Biggl|_{\rho^{\;'}=a}=-{\cal F}
\left[f_n(k)\,e^{\imath\left(k(z{\;'}-z^{0})+n\vfi{\;'}\right)}
\right];\;\;
d\,\frac{\partial\Fip_\mu}{\partial \rho^{\;'}}\Biggl|_{\rho^{\;'}=a}={\cal F}
\left[
c_\mu^\pm f_{n\pm\,1}(k)\,e^{\imath\left(k(z{\;'}-z^{0})+n\vfi{\;'}\right)}
\right]
\label{DFip0}\\
d\,\frac{\partial\Fip_{\mu\nu}(\vec{r}\;^{'})}{\partial \rho^{\;'}}\Biggl|_{\rho^{\;'}=a}=-{\cal F}
\left[
c_\mu^\pm\,c_\nu^\pm f_{n\pm\,2}(k)+\left(\delta_{\mu\,\nu}/2\right)f_n(k)
\right]
e^{\imath\left(k(z{\;'}-z^{0})+n\vfi{\;'}\right)} \;,\qquad\qquad\label{DFipmunu}
\eea
with the same notations and same order of the remainder as in the formulas (\ref{uDFip0}), (\ref{uDFipmunu}), assuming again condition (\ref{newcond}).
\vfill\eject

\vfill\eject

\end{document}